# Highly efficient, dual state emission from an organic semiconductor

Sebastian Reineke[1,2,a)], Nico Seidler[3], Shane R. Yost[1,4], Ferry Prins[1,5], William A. Tisdale[1,5], and Marc A. Baldo[1,2].

[1] Energy Research Frontier Center for Excitonics, Massachusetts Institute of Technology, 77 Massachusetts Avenue, Cambridge, MA 02139, USA.

[2] Department of Electrical Engineering and Computer Science, Massachusetts Institute of Technology, 77 Massachusetts Avenue, Cambridge, MA 02139, USA.

[3] Department of Physics and Astronomy and London Centre for Nanotechnology, University College London, Gower Street, London WC1E 6BT.

[4] Department of Chemistry, Massachusetts Institute of Technology, 77 Massachusetts Avenue, Cambridge, MA 02139, USA.

[5] Department of Chemical Engineering, Massachusetts Institute of Technology, 77 Massachusetts Avenue, Cambridge, MA 02139, USA.

We report highly efficient, simultaneous fluorescence and phosphorescence (74% yield) at room temperature from a single molecule ensemble of (BzP)PB [N,N'-bis(4-benzoyl-phenyl)-N,N'-diphenyl-benzidine] dispersed into a polymer host. The slow phosphorescence (208 ms lifetime) is very efficient (50%) at room temperature and only possible because the non-radiative rate for the triplet state is extremely low ($2.4 \times 10^0$ s$^{-1}$). The ability of an organic molecule to function as an efficient dual state emitter at room temperature is unusual and opens new fields of applications including the use as broadband down-conversion emitters, optical sensors and

a) Author to whom correspondence should be addressed. Electronic mail: reineke@mit.edu.

attenuators, exciton probes, and spin-independent intermediates for Förster resonant energy transfer.

**MAIN TEXT**

Unlike other luminescent, solid-state systems such as inorganic semiconductors and quantum dots,[1] organic semiconductors are characterized by highly localized excitons with strong exchange interactions. Localization is arguably the principal advantage of organic materials in displays. It helps protect the efficiency of light emission even in the highly-disordered thin films that are compatible with cost-effective manufacturing. But localization in organic semiconductors also generates distinct excited states, defined by the large exchange splitting between the spin 0 and 1 excited states, known as the singlet ($S_1$) and triplet ($T_1$), respectively, and having vastly different properties.[2] Efficient photoluminescence (fluorescence) can occur from the singlet state where the relaxation conserves spin. Luminescence from the triplet state (phosphorescence) is, however, typically very weak because it involves a spin flip, which is quantum mechanically forbidden.[2,3] Thus, triplet states in organic semiconductors are typically dark states.

In electronics, about 75% of injected charges form dark triplet states. Consequently, triplet state engineering is a distinctive and crucial problem in organic semiconductors. Triplets are present in the vast majority of organic-based optoelectronic applications including lasers[4], photovoltaic cells[5], and light-emitting diodes[6,7]. The problem of dark triplet states has motivated the use of organometallic complexes[6] that enable efficient phosphorescence via strong spin orbit coupling and, more recently, thermally activated delayed fluorescence[8], which minimizes the exchange splitting, enabling effective reverse intersystem crossing from the triplet back to the luminescent singlet.

Few, if any, purely organic materials have demonstrated biluminescence - efficient light emission from *both* the singlet and triplet states. This is explained in Figure 1, which illustrates the energy state diagram of an organic lumophore with its lowest singlet ($S_1$) and triplet ($T_1$) states, which are split by the exchange interaction $\Delta E_{ST}$ (Ref. 2,3). The photoluminescence (PL) quantum efficiency of biluminescence $\eta_{BL}$ is given by Equation (1):

$$\eta_{BL} = \Phi_F(1 - \Phi_{ISC})\left[\frac{1}{1-\Phi_{ISC}\Phi_{RISC}}\right] + \Phi_P\Phi_{ISC}\left[\frac{1-\Phi_{RISC}}{1-\Phi_{ISC}\Phi_{RISC}}\right]. \qquad (1),$$

Here, $\Phi_F$ and $\Phi_P$ – the respective quantum yields of fluorescence and phosphorescence – are defined as the ratio of the radiative over the sum of radiative and non-radiative rates of the respective state [e.g. $\Phi_F = k_{r,F}/(k_{nr,F} + k_{r,F})$]. The yield of intersystem crossing from $S_1$ to $T_1$ is $\Phi_{ISC} = k_{ISC}/(k_{nr,F} + k_{r,F} + k_{ISC})$ and reverse intersystem crossing from $T_1$ to $S_1$ is $\Phi_{RISC} = k_{RISC}/(k_{nr,P} + k_{r,P} + k_{RISC})$. Conventional fluorophores like laser dyes have $\Phi_{ISC} \approx 0$, eliminating the possibility of phosphorescence from the triplet. In such materials with weak spin orbit coupling, the phosphorescence radiative rate can be as slow as $k_{r,P} \sim 10^0$ s$^{-1}$, which typically is outcompeted by an orders of magnitude faster non-radiative rate, i.e. $k_{nr,P} \sim 10^2 - 10^6$ s$^{-1}$ (Ref. 3). Consequently, the intersystem crossing in most purely organic semiconductors only represents a loss channel. On the contrary, highly efficient phosphorescent molecules require strong spin orbit coupling to mix the singlet and triplet states, and consequently, they possess $\Phi_{ISC} \approx 1$, eliminating the possibility of fluorescence from the singlet. Rather, the key to biluminescence is moderate values of $\Phi_{ISC}$, which can only be obtained with weak spin orbit coupling. Thus, the phosphorescent radiative rate in a biluminescent material is necessarily very slow, and obtaining efficient phosphorescence is challenging.

There are few observations of phosphorescence of purely organic molecules at room temperature: in silica glasses ($\Phi_P$ = n.a.)[9], in hydrogen-bonded polymer matrices ($\Phi_P$ = n.a.)[10], and of graphene quantum dots (weakly biluminescent, $\eta_{BL}$ = 0.02)[11], which all require very specific fabrication procedures. Recently, persistent phosphorescence from organic materials – even under ambient conditions using novel host systems – has been demonstrated ($\Phi_P\Phi_{ISC}$~10%), where the non-radiative deactivation has been minimized through deuterization of the organic molecules.[12] In an ideal biluminescent system with $\Phi_F$ and $\Phi_P$ approaching unity, intersystem crossing does not introduce loss to the system but only mixes the relative intensities of fluorescence and phosphorescence.

In this letter, we discuss the luminescent properties of (BzP)PB [N,N'-bis(4-benzoyl-phenyl)-N,N'-diphenyl-benzidine] (chemical structure shown in Figure 1), a purely organic small molecule, which shows intense emission from both singlet and triplet at room temperature, fulfilling the requirements for biluminescence. For our study, we have chosen the conventional, inert polymer PMMA [Poly(methyl methacrylate)] as wide band gap host material ($T_1$ = 3.1 eV)[13] based on its ability to form rigid glasses, which is beneficial for observation of phosphorescence at room temperature.[14] Substrates were cleaned subsequently through a 5 min ultrasonic bath in acetone and a 2 min immersion (boiling) in isopropyl alcohol. 2 wt% of the organic dye with respect to the polymer is spun down from a methoxybenzene solution in a protected nitrogen atmosphere onto quartz substrates (1" by 1" by 1mm) at a spin speed of 2000 rpm (1000 rpm/s ramp) for 60 s. Materials were used as received from the vendor without any purification steps: methoxybenzene from Sigma Aldrich; PMMA from Alfa Aesar; and (BzP)PB from Sensient Imaging Technologies GmbH. Photoluminescence (PL) spectra were recorded with a radiometrically-calibrated spectrometer (USB2000, Ocean Optics Inc.). The setup for the transient PL consisted of a 365 nm light-emitting diode as excitation source driven by a LEDD1B driver (both Thorlabs), an amplified silicon detector (PDA36A, Thorlabs) as detector, and an oscilloscope for data collection (Tektronix TDS 3054C). The synchronization of pulse train and detection was achieved using a pulse generator with optional delay (Hewlett Packard B114A). Delayed spectra were collected in a similar fashion, using the spectrometer as detector in a time-gated scheme. Here, the time offset between pump off and detection start were kept short (< 500 µs) to detect virtually all delayed emission, but long enough to filter out prompt fluorescence. The fluorescence dynamics measurements were performed using an inverted microscope (Nikon Ti Eclipse) equipped with a 20× objective (N.A. = 0.4). The sample was excited using a pulsed 405 nm laser diode (LDH-D-C-405M, Picoquant, 20 MHz repetition rate) with an average fluence of 60 nJ / $cm^2$ across a spot size of ~2 µm in diameter. The sample luminescence was detected using a Si avalanche photodiode (PDM50, Micro Photon Devices), after filtering out the excitation using a set of fluorescent filters (Semrock). Data acquisition was performed using a time correlated single photon counting board (PicoHarp 300, Picoquant). The full-width-at-half-maximum (FWHM) of the instrument response function (IRF) was 0.42 ns. Measurements at 77 K were carried out in a bath of liquid nitrogen, whereas all remaining photoluminescence (PL) experiments were carried out in a nitrogen glovebox with ~2ppm oxygen and <2ppm water content. To align 77 and 293 K data for relative intensities, the sample's integrated PL was obtained at both temperatures in a cryostat (Advanced Research Systems, Inc.) without changing the geometry. PL quantum yield measurements were performed in a 6" integrating sphere (Labsphere) using a 405 nm laser (CPS405, Thorlabs) as excitation source and the USB2000 spectrometer, following known protocols[15].

In order to discriminate between the emission from singlet and triplet states, we have performed time-gated spectroscopy. The PL spectra of (BzP)PB are shown in Figure 2, all collected at room temperature. The solid spectrum shows the integrated spectrum of 2 wt% (BzP)PB doped in a PMMA host polymer obtained under cw-illumination. The dashed-dotted spectrum is the total emission collected after turning off the pump source (UV LED). It is a mixture of delayed fluorescence (DF), observed as a result of a small singlet-triplet splitting $\Delta E_{ST}$ = 290 meV (cf. Ref. 8), and phosphorescence (P). By subtracting the total delayed emission from the integrated, steady state emission, we obtain the prompt fluorescence, shown in Figure 2 as dashed line. Fitting the spectral distribution of the fluorescence to the total delayed spectrum and subtracting it gives the phosphorescence spectrum (dotted line).

Figure 3 shows the PL dynamics of the PMMA:(BzP)PB 2 wt% thin film at room temperature. In Figure 3 (a), the prompt fluorescence decay following a sub-nanosecond pulsed excitation at 405 nm is plotted. The fluorescence shows a bi-exponential decay with respective lifetimes of 2.0 and 5.1 ns. In addition to the prompt singlet emission, the transient shows a third component, which can be treated as a constant background in the time window shown. It is the contribution of the delayed emission with origin in the triplet manifold (both DF and P). The onset of this delayed emission is determined by the long lifetime of the triplets (see below) and the repetition rate of the laser source (20 MHz). Figure 3 (b) shows the transient PL signal of the delayed emission, as obtained directly after the 120 ms, quasi-cw 365 nm pump pulse. The decay is plotted for both room temperature (293 K) and 77 K. The transients show two time constants with the faster being DF (40 ms at 293 K) and the slower P (208 ms at 293 K). The dashed and dashed-dotted lines in Figure 3 (b) indicate the phosphorescence contribution at the respective temperature. At room temperature, 7.4% of the total cw intensity is phosphorescence. Because the initial curvature of the transient does not depend on pump intensity, DF is only a result of thermal activation[8] and not of triplet-triplet annihilation[16,17]. As expected for a thermally assisted process, the time constant of DF increases to 60 ms at 77 K and its overall contribution is reduced by a factor of 5.3. Interestingly, the overall phosphorescence intensity is 1.8 times higher at room temperature compared to the value at 77 K. This behavior is unusual, as long-lived phosphorescence in organic molecules readily decreases with increasing temperature, as non-radiative modes strongly increase.[12] This suggests that intersystem crossing might be thermally assisted, and that the coupling of the (BzP)PB triplet state to intramolecular rotational and vibrational as well as host-guest phonon modes is exceptionally weak.

The PL quantum yield $\eta_{BL}$ of the PMMA:(BzP)PB 2 wt% thin film is 74% at room temperature under cw illumination. However, a higher quantum yield (93%) is observed for pump durations shorter than the phosphorescence lifetime. Figure 4 shows PL transients of the sample under different pump source widths spanning from 1 to 500 ms. As a result of the long triplet lifetime (208 ms), the phosphorescence slowly ramps up in intensity. For long pulse widths, the biluminescent intensity during the pulse is reduced reaching a steady state at 80% of the initial intensity. For all pulse widths, the sample absorption during the pump duration is constant with time. The decrease is observed in the prompt fluorescence and is a result of increasing singlet re-absorption in the triplet manifold ($k_{SA}$ in Figure 1) upon the build-up of the $T_1$ population. Consequently, one can deduce the quantum efficiency in the absence of a significant triplet density to be 93%, which also marks a lower limit for the singlet state efficiency $\Phi_F \leq 93\%$, as we are unable to differentiate between $k_{nr,F}$ and $k_{ISC}$. Considering the reduction of $\eta_{BL}$ from 93 to 74% to be a loss in the singlet manifold due to $k_{SA}$ and a 7.4% share of phosphorescence, the triplet state efficiency $\Phi_P$ can be estimated to be at least 50%. Taking the triplet radiative lifetime of (BzP)PB (208 ms) into account, the corresponding triplet state non-radiative rate in the present system has a very low value of $k_{\text{nr,P}} = 2.4 \cdot 10^0 \text{ s}^{-1}$. Thus, (BzP)PB – a purely organic small molecule – emits very efficiently from both its singlet and triplet state. With non-radiative rates for both spin species approaching zero, the luminescent properties of (BzP)PB are determined solely by the mixing of singlets and triplets through $k_{\text{ISC}}$ and $k_{\text{RISC}}$ (cf. Figure 1 and Equation (1)). This mixing is set by the molecular structure[12] and depends on the electronic and spin orbit coupling between the two states as well as on their energetic splitting $\Delta E_{\text{ST}}$ (Ref. 3).

In summary, we have demonstrated that highly efficient, simultaneous fluorescence and phosphorescence can coexist in a single organic small-molecule material at room temperature. To enable efficient phosphorescence, the molecules were diluted into a rigid polymer host. Biluminescence, spanning the wide, virtually unexplored field between strictly fluorescence and purely phosphorescence emitting molecules, holds great promise for novel applications. The observation of the low non-radiative rate ($\sim 10^0$ s$^{-1}$) found in these organic molecules suggests that efficient dual state emission with various transition strengths in the triplet manifold can be customized in single molecule probes of exciton spin dynamics.

**ACKNOWLEDGEMENTS**

This work was supported as part of the Center for Excitonics, an Energy Frontier Research Center funded by the U.S. Department of Energy, Office of Science, Office of Basic Energy Sciences under Award Number DE-SC0001088 (MIT). SR gratefully acknowledges the support of the Deutsche Forschungsgemeinschaft through a research fellowship (Grant No. RE3198/1-1).

**FIGURES**

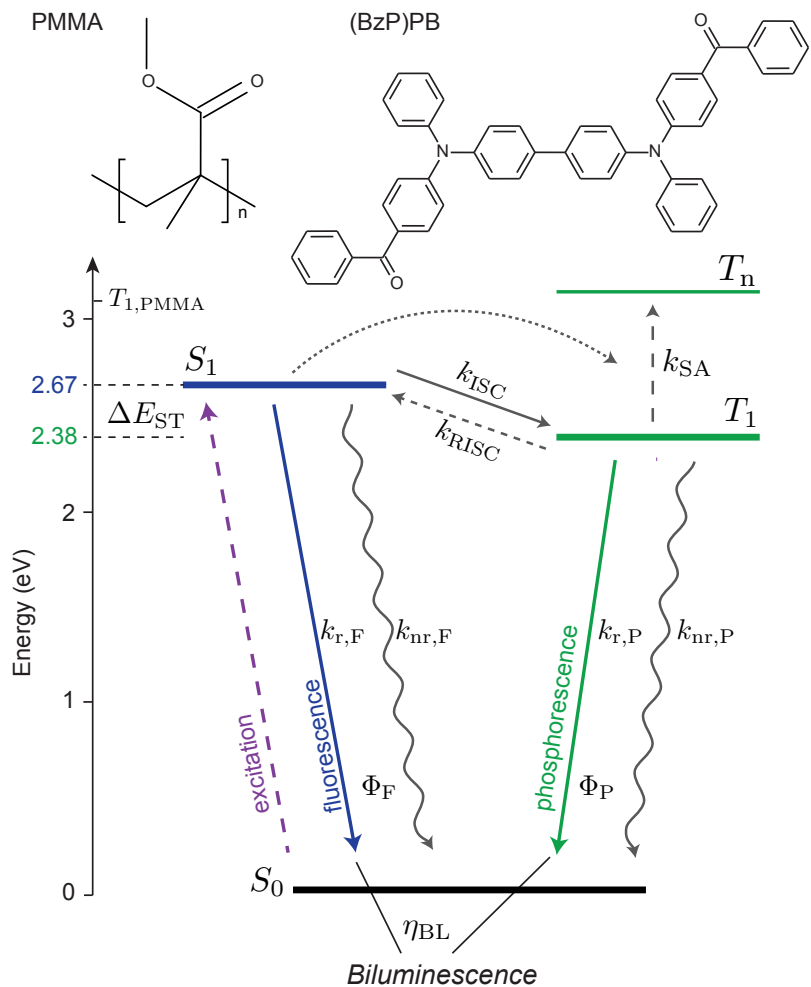

FIG. 1. (color online) Energy level diagram indicating all possible rates $k_i$ of (BzP)PB, a biluminescent emitter, showing emission from singlet $S_1$ and triplet $T_1$ states. Chemical structures of PMMA host polymer and (BzP)PB are shown. The subscripts 'F' and 'P' refer to fluorescence and phosphorescence. Further, 'ISC' and 'RISC' are intersystem and reverse intersystem crossing, 'r' and 'nr' radiative and non-radiative, $\Phi_i$ the individual state efficiencies; 'SA' denotes singlet absorption in the excited triplet manifold, $\Delta E_{ST}$ is the singlet-triplet splitting, and $\eta_{BL}$ the integrated emission quantum efficiency.

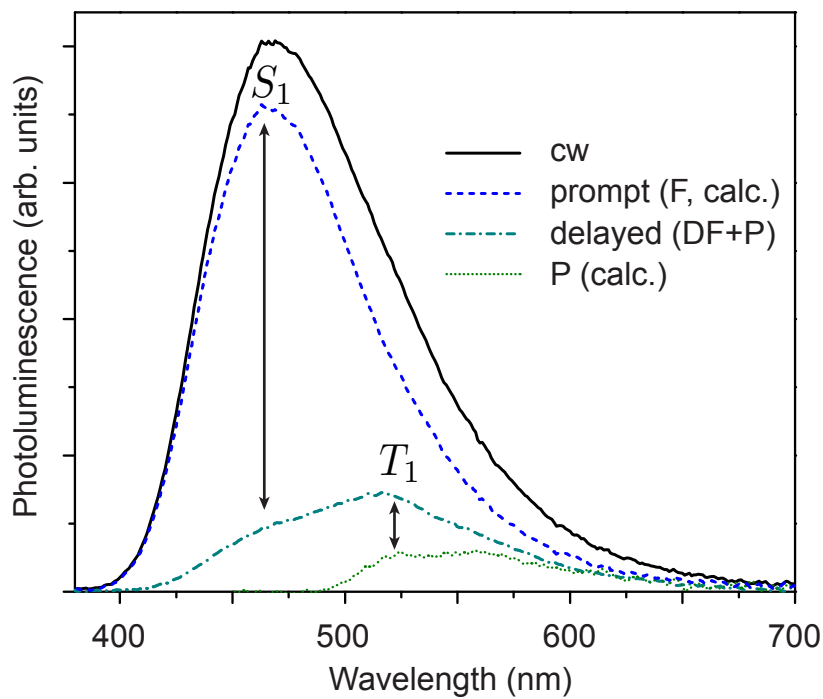

FIG. 2. (color online) Photoluminescence spectra of (BzP)PB at room temperature (293 K). Black line is the integrated emission obtained cw-illumination. Dash-dotted line represents the total delayed emission consisting of delayed fluorescence ('DF') and phosphorescence ('P'). Prompt F (dashed) and P (dotted) are obtained from integrated and delayed intensity through consecutive processing steps (see text). Arrows indicate singlet $S_1$ and triplet $T_1$ state.

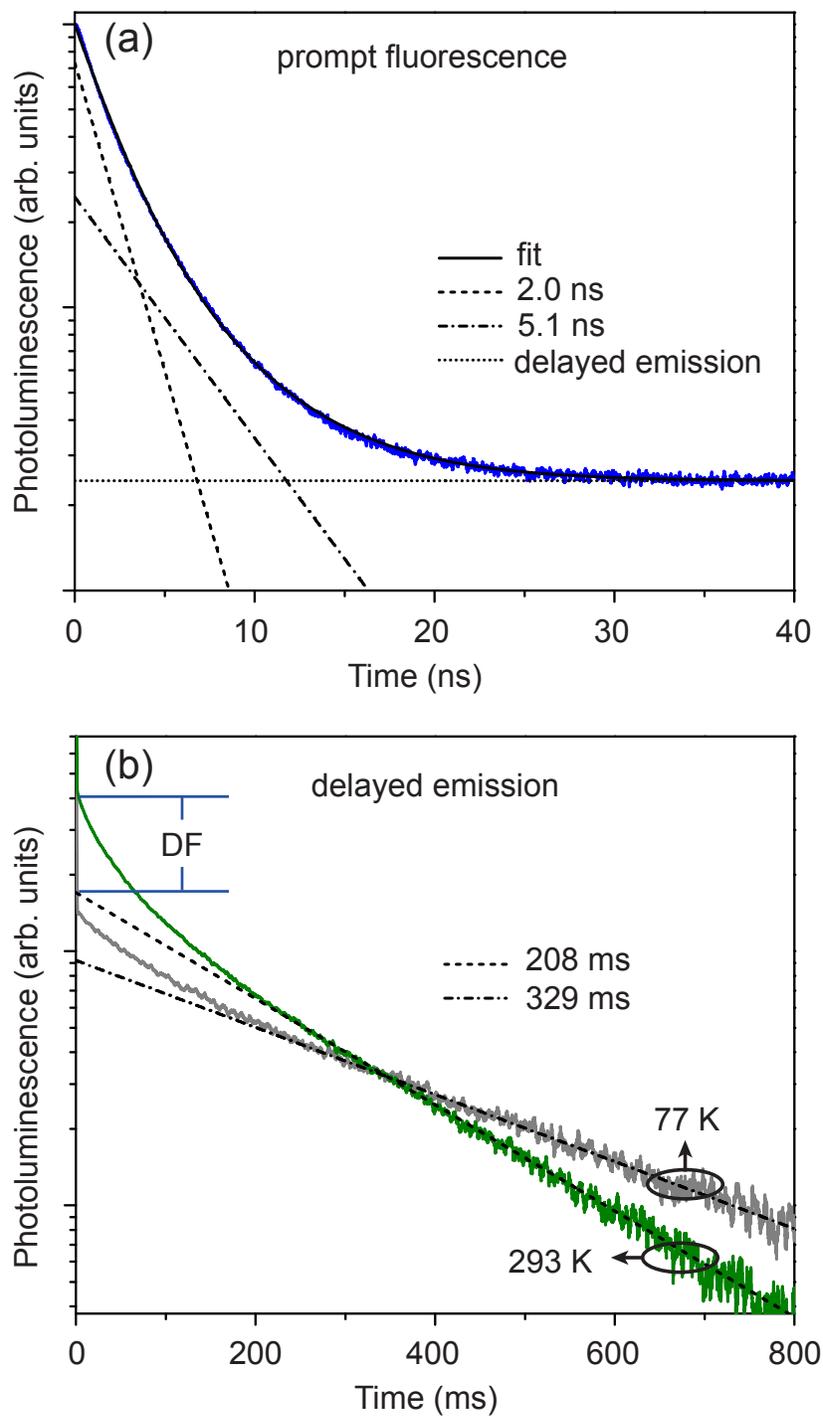

FIG. 3. (color online) Time-resolved photoluminescence of (BzP)PB. (a) Prompt fluorescence: solid line is a calculated fit based on a bi-exponential decay plus constant background (dotted line) accounting for the delayed emission. Dashed and dash-dotted lines indicate the components

of the prompt fluorescence. (b) Delayed emission: Transient signals following a 120 ms, quasi-cw pump pulse for 293 and 77 K. Dashed (293 K) and dash-dotted (77 K) lines indicate the phosphorescence contribution. The initial faster decay is delayed fluorescence ('DF').

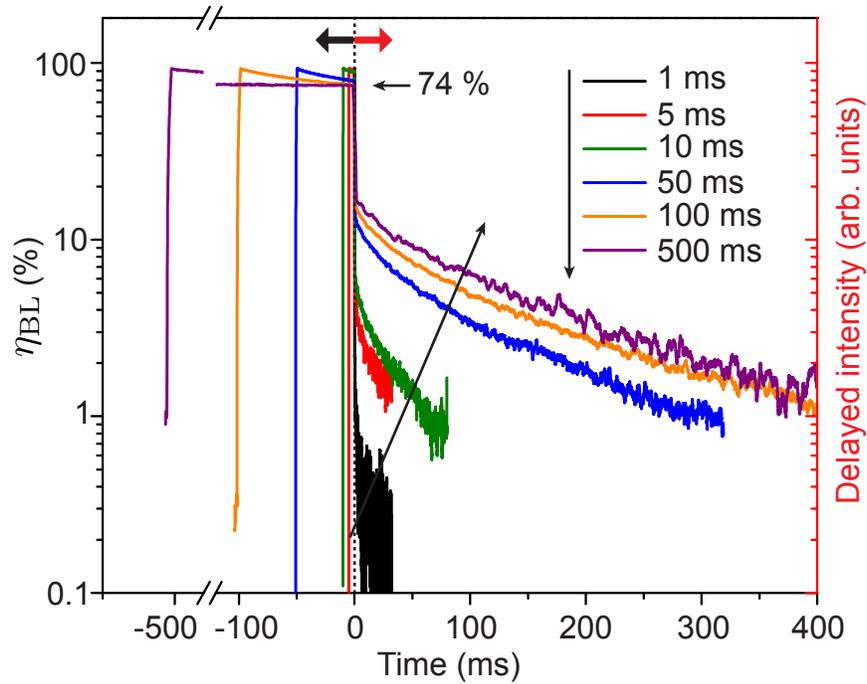

FIG. 4. (color online) Pump duration dependency of photoluminescence quantum yield $\eta_{BL}$ (to the left of the vertical dashed line at $t = 0$) and triplet population (to the right). Arrows are indicating increasing pump duration. Maximum $\eta_{BL}$ value obtained for short durations and at the beginning of the long pulses is 93%. The steady state $\eta_{BL}$ is 74%.